\documentclass[a4paper]{spie}  
\usepackage{amsmath}
\usepackage{mathabx}
\usepackage[]{graphicx}
\usepackage{rotating,tabularx}
\usepackage{textcomp}

\def \hfillx {\hspace*{-\textwidth} \hfill}

\title{RAFTER: Ring Astrometric Field Telescope \\ 
    for Exo-planets and Relativity } 

\author{A. Riva\supit{a}, M. Gai\supit{a}, A. Vecchiato\supit{a}, D.  Busonero\supit{a}, M. G. Lattanzi\supit{a}, F. Landini\supit{a}, Z. Qi\supit{b},  Z. Tang\supit{b}
\skiplinehalf
\supit{a}Istituto Nazionale di Astrofisica - Osservatorio Astrofisico di Torino, V. Osservatorio, 20, I-10025 Pino Torinese (TO), Italy \\
\supit{b}Shanghai Astronomical Observatory, Chinese Academy of Sciences, 80 Nandan Road, Shanghai 200030, China\\
}
\authorinfo{Further author information: E-mail: alberto.riva@inaf.it, Telephone: +39 011 8101 963, www.oato.inaf.it}

\begin{document} 
  \maketitle 

\begin{abstract}

High precision astrometry aims at source position determination to a very 
small fraction of the diffraction image size, in high SNR regime. 
One of the key limitations to such goal is the optical response variation 
of the telescope over a sizeable FOV, required to ensure bright 
reference objects to any selected target. 
The issue translates into severe calibration constraints, and/or the need 
for complex telescope and focal plane metrology. 
We propose an innovative system approach derived from the established TMA 
telescope concept, extended to achieve high filling factor of an annular 
field of view around the optical axis of the telescope. 
The proposed design is a very compact, 1 m class telescope compatible with 
modern CCD and CMOS detectors (EFL = 15 m). 
We describe the concept implementation guidelines and the optical performance 
of the current optical design. The diffraction limited FOV exceeds 1.25 square 
degrees, and the detector occupies the best 0.25 square degree with 66 devices.

\end{abstract}



\section{Introduction}
\label{sec:introduction}  

Astrometric measurements at or below the micro-arcsec ($\mu as$) level 
appear to be compatible with even moderate values of telescope diameter 
and exposure time, for observation of fairly bright stars in a wide spectral 
band in the visible range (as recalled below). 
We take as reference science case the search for Earth-like exo-planets in the 
population of nearby (bright) stars with spectral type not too different from 
our Sun. 

Astrometry \cite{Shao19}, by measurement of the reflex motion of parent 
stars on the sky, is expected to be able to allow determination of 
individual planet’s mass and orbital inclination, and ease the full solution 
of multiple systems. 
Competing techniques, as radial velocity and transits, are most sensitive to 
specific target geometries and/or astrophysical characteristics, so that most 
currently detected exo-planetary systems are quite different from our solar 
system. 

This consideration motivates the proposition in recent years of projects 
like NEAT \cite{Malbet12}, Theia \cite{Malbet16}, TOLIMAN 
\cite{Bendek18} and others \cite{Hahn18}, devoted to astrometric 
surveys e.g. of the limited sample of several ten nearest and 
brightest near-solar stars, in particular on already detected systems, 
in which $\sim 1\,\mu as$ measurement precision may allow accurate 
determination of Earth's analogues. 
We recall that for a $\sim 1\, M_\Earth$ planet at $\sim 1\, AU$ from a near-solar 
star, i.e. within its habitable zone (HZ), the astrometric signal is 
$\sim 0.3 \, \mu as$. 
Some of the proposers are also engaged in the bilateral (China-Italy) research 
project ASTRA (Astrometric Science and Technology Roadmap for Astrophysics)\cite{Gai2020ASTRA}, 
presented in another contribution to these Proceedings. 

The limiting astrometric precision\cite{Lindegren1978, GaiPASP1998, Mendez2013} 
is associated to the instrument resolution and the photon budget. 
However, preserving a comparable measurement accuracy on the angular separation 
among stars is not so easy, since the realistic optical response of a system 
implies a Point Spread Function (PSF) variation over the field, due to 
aberrations, which affects the photo-centre determination. 
Also, the photons from different viewing directions, associated to different 
focal plane positions, use different parts of the optical system; the 
non-common parts of the optical paths are affected by further variations 
(``beam walk'') due not only to the field dependence of aberrations, but 
also to the different characteristics of different areas of the optical 
surfaces, e.g. from figuring and micro-roughness. 
Actually, the knowledge on instrument response, by calibration and metrology, 
mitigates the problem, so that ``only'' residual effects are to be taken 
into account; unfortunately, they are still significant  at the 
$\mu as$ level. 

It should be noted that high precision astrometry implicitly requires 
a correct implementation of the measurement framework in the context 
of General Relativity (GR). 
Gravitational deflection of the photons \cite{Will2006} 
by the space-time curvature induced by the Sun alone corresponds to 
several milli-arcsec (hereafter, $mas$) over most of the sky accessible 
by an observer in the vicinity of the Earth. 
In many cases, the effects of Solar system planets not far away from the 
observing direction may be significant at the (sub)-$\mu as$ level. 

The real challenges to accuracy stem from three main potential sources of 
systematic errors: 

\begin{itemize}
\item Variable optical response of the telescope; 
\item Variable electro-optical response of detector; 
\item ``Cosmic noise", i.e. intrinsic variability of astronomical objects; 
\end{itemize}

The annular field telescope concept arises mainly as a possible approach 
toward answer to the first issue, although some considerations about the 
others will be proposed as well. 

Optical response is of course one of the main design drivers, since usually 
science requirements are translated into specifications for image quality over 
a suitable field of view (FOV). 
The need for a minimum size of the instrument FOV 
is related to assuring the capability of observing several sources 
simultaneously, for reasons of efficiency and to provide references 
to the science target (calibration). 
Besides, the instrument response typically changes over a large FOV, 
both in terms of optical parameters and of detector characteristics, 
in turn affecting the calibration requirements.

Therefore, the requirement of good optical response, possibly
uniform, over the whole field of view, is a system level challenge, from  
optical design, through manufacturing and testing, to integration
and in-flight alignment. 
It remains a critical aspect throughout science operation, because 
instrument calibration must be maintained over the useful lifetime. 

\subsection{Rationale for an annular field telescope}
\label{sec:rationale}
In order to alleviate some of the above challenges, we investigate a 
novel instrument concept, in which a telescope is optimized 
on an annular region set at a given radius from the optical axis. 
In a nominal optical system characterized by circular symmetry, the 
optical response is also the same, apart rotation, over any concentric 
ring of the focal plane. 
We remark that the ring region corresponds to a small set of angular 
values, just replicated by symmetry. 
The annular field can provide simultaneous observation of any target 
pair with separation ranging from zero to its angular diameter. 
On the other hand, the annular field may not be convenient for some 
applications, e.g. to perform a uniform scan of a large area. 

The optical design is based on the same optimization tools used to
achieve conventional large field telescopes already presented 
in the literature, just by reformulation of the cost function, 
so that, upon convergence, the optical response can be expected 
to be as good, uniform and reliable as other configurations 
analyzed. 
We are investigating the feasibility of such design approach. 

An application of the annular field telescope concept is 
presented in another contribution to these Proceedings, describing 
two configuration options for the AGP (Astrometric Gravitation Probe) \cite{Gai2020AGP}
mission concept. 

\begin{figure}
\centering
\includegraphics[width=0.33\textwidth]{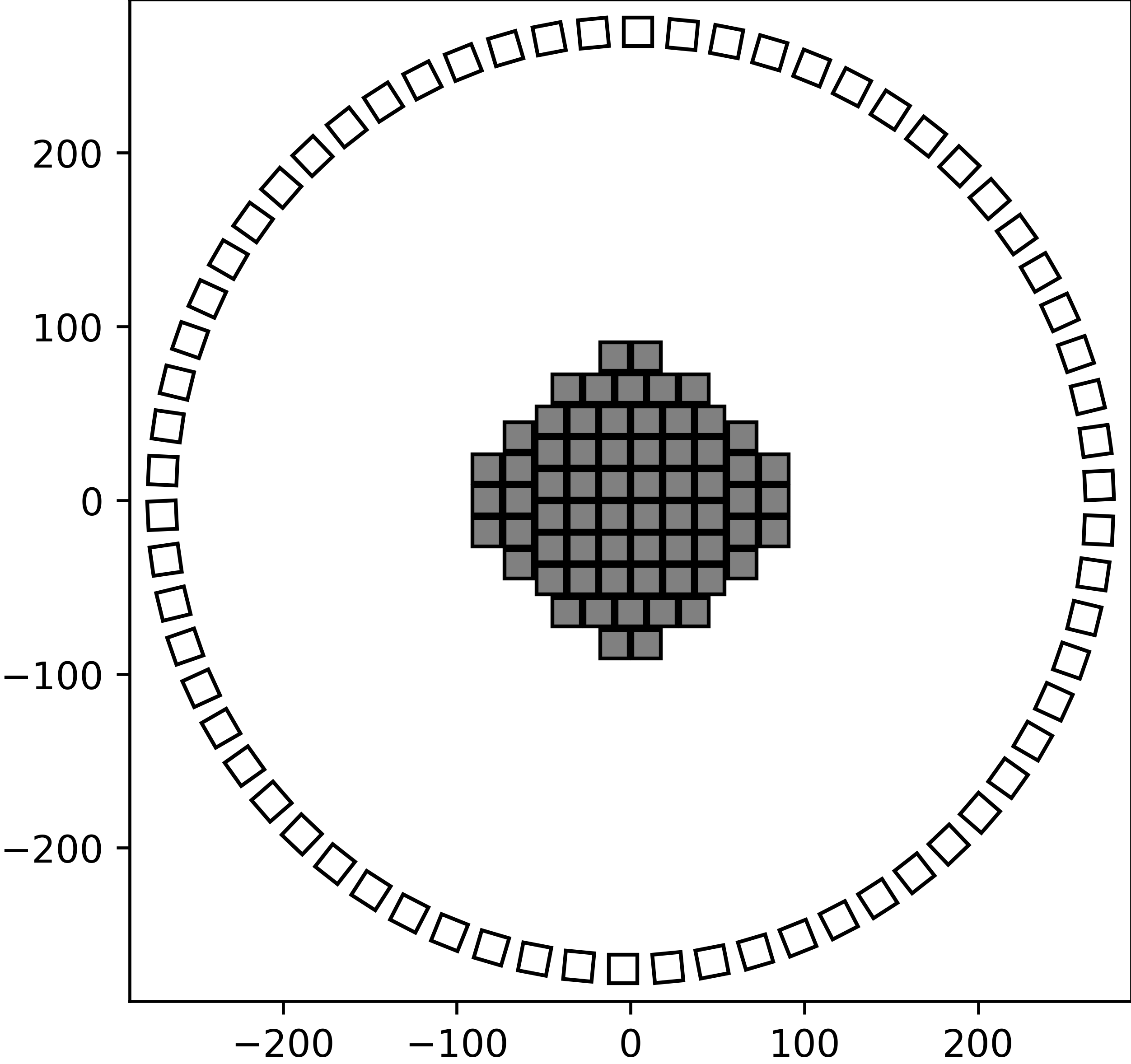} 
\caption{\label{fig:FPRing} Ring focal plane with radius 
$1^{\circ}$, composed of 66 detectors, and the same 
devices (in gray) arranged around the optical axis. 
Plot axes are in mm. }
\end{figure}

An annular system with detectors covering the limited radial range 
between $\theta$ and $\theta + \delta\theta$ still provides a significant
total area $2\pi\theta \cdot \delta\theta$, but much less than that
of the inscribed circle $\pi\theta^{2}$ when the ring is large with
respect to its thickness, i.e. $\delta\theta << \theta$. 
An example of annular field, with radius 
$\theta = 1^\circ$, is shown in Fig.~\ref{fig:FPRing}; 
the ring width is $\delta \theta = 3'.75$. 
The same number of detectors used in the ring are also shown (in gray) 
in a compact distribution around the FOV center, for comparison with 
a conventional mosaic having comparable complexity. 
The total field of view is $0.258\,$square 
degrees, covered by 66 detectors, but in the conventional case 
(filled central area, gray in figure) the maximum angular separation 
between simultaneously imaged sources is much smaller, i.e. 
$\sim 37\, arcmin$ against $2^\circ$. 

It may be noted that any instrument aiming at filling one square degree 
with $\sim 50 \, mas$ pixels is fated to end up with a pixel count of 
$\sim \pi \left( 20 \times 3600 \right)^2 \simeq 1.6 \times 10^{10}$ 
pixels, i.e. $\sim 1,000$ individual $4k \times 4k$ chips. 
Availability of larger format devices would alleviate some of the 
system complexity, but not the total pixel count. 
\\ 
Even if negligible variation of the optical response over the focal 
plane were achieved, which is one of our key requirements, the sheer cost 
of such detection system is dramatically in favour of our alternative 
annular field approach. 

The key point addressed in this paper is the feasibility and potential 
performance of the annular field approach, in particular with respect 
to optical design and system considerations. 
In Sec.\,\ref{sec:OptDesign} we describe the design development guidelines 
and main results on optical performance; in Sec.\,\ref{sec:Discussion} 
we briefly review the highlights of the proposed concept, in particular 
concerning astrometric capabilities; finally, in Sec.\,\ref{sec:conclusions}, 
we draw our conclusions outlining future investigations. 

\section{Concept development and Optical design}
\label{sec:OptDesign} 
The concepts investigated in the framework of NEAT \cite{Malbet12} and 
Theia \cite{Malbet16} (proposed respectively in response to ESA's M3 and 
M4 Calls) address the astrometric accuracy issues by selecting either a 
very simple telescope (one mirror), or a high performance system (three 
mirror anastigmat, TMA), providing a good optical response, uniform over 
a convenient FOV. 
Moreover, the instrument design was endowed with advanced metrology and 
calibration systems to monitor the astrometric response and allow correction 
of the still existing systematic errors in the data reduction. 

In this paper, we build on similar concepts, exploiting the symmetry
of an optical system derived from a TMA design, and introducing an
optimization approach restricted to an annular area of the focal plane.
A ring centered on the optical axis of a nominal on-axis system is
basically considered as a single point with respect to optical quality,
just replicated by circular symmetry over the selected circle. A narrow
annular region can therefore be expected to feature a small radial
variation in optical response, with very uniform characteristics in
the  azimuthal direction. 

The optical design of large aperture, large field of view telescopes 
with three or four mirrors endowed with optical power, plus additional 
folding mirrors, has been consolidated in the past few decades, in 
particular with the concept of Three-Mirror Anastigmats (TMA) developed 
by D. Korsch \cite{Korsch77,Korsch80}. 
TMA telescopes have been adopted for important science experiments, 
e.g. the Euclid mission \cite{Euclid2016b} and the James Webb Space 
Telescope (JWST) \cite{Contreras04,Davila04}. 
The GAME \cite{GameExpAstr2012,Gai_GAME_2014} and AGP 
\cite{AGP_Riva16,AGP_Landini16} concepts (submitted respectively 
to the M3 and M4 ESA Calls) also rely on a TMA derived telescope, 
enforcing the concepts of differential measurement of fields close 
to the Sun, affected by fairly large light deflection, with a 
dedicated design deployed around a coronagraphic subsystem. 

TMAs may achieve extremely good performance, but they also feature 
comparably high sensitivity to misalignment \cite{Thompson08}, 
as understandable from their large \'{e}tendue (product of aperture 
area and field angular coverage). 
The annular field telescope proposed in this paper actually achieves an 
{\em intermediate} \'{e}tendue, since the field of view range is large, 
but not filled, i.e. only a fraction of the subtended phase space 
is actually used. 

A typical TMA feature is a significant vignetting of the central 
region, so that they are often used in an off-axis configuration; e.g., 
JWST is on-axis in aperture and off-axis in field. 
Our proposed annular field can be considered as an off-axis case, 
replicated by symmetry around the optical axis (resulting in an on-axis telescope that uses only a corona region of focal plane); ring field 
telescopes have been proposed e.g. for Earth observation 
\cite{Calamai15}, with a thorough discussion of the design 
approach and performance. 
Actually, an astronomical telescope with annular field was proposed for 
the DUNE concept \cite{Grange06}, but this design feature aspect was dropped
when it was merged with the SPACE proposal in the Euclid space mission 
\cite{Content08}. 

\subsection{ Science specifications \label{sec:TechSpec}}
The main performance driver of our study is astrometric accuracy in the 
determination of angular distance between bright, unresolved sky objects. 
We set our reference magnitude in the range $8-10\,mag$ at visible wavelengths, 
for near-solar type stars. 
The observing wavelength is set in the visible: $\lambda \simeq 0.55\,\mu m$, 
with a fairly large spectral bandwidth $\Delta \lambda \simeq 200\,nm$.  
The number of photons per hour received from a $D=1\,m$ diameter telescope, 
with a $0.6\,m$ central obstruction (case explored in our study, see below), 
ranges between $\sim 3.6e9$ ($10\,mag$) to $\sim 2.3e10$ ($8\,mag$). 

This is expected to be compatible with the precision goal, assuming repeated 
visits to each target throughout a mission lifetime of $\sim 5$\,years, and 
a $1\,m$ class telescope operating close to the diffraction limit. 
The angular resolution, corresponding to the characteristic image size, 
is $\sim \lambda/D\simeq110\,mas$. 

The photon limited precision on location in 1\,hour of observation, in each 
coordinate, is $\sim 2\,\mu as$ for a $10\,mag$ unresolved source. 
In any case, the observation is assumed to be split in a sequence of shorter 
exposures, limited by detector saturation to a few or several 
$10^4$\,photo-electrons. 
This corresponds to an elementary exposure time of order of a few $10\,ms$, 
feasible by acquisition of selected regions of interest, if not by full frame 
readout. 

Astrometric uncertainty scales with photons to $\sim 0.7\,\mu as$ (8\,mag); 
moreover, the nominal location dispersion 
(and similarly the precision on related parameters, e.g. parallax or orbit 
semi-axis) scales with the square root of exposure time (e.g. by a factor 
4.9 in 24 hours) and of the number of visits $N_V$, so that sub-$\mu as$ 
performance is achievable, in an ideal world. 
In practice, it is usually considered unavoidable to hit the ``floor" 
imposed by instrument perturbations and, ultimately, calibration 
uncertainties. 

One of the main drivers of our study is the definition of a measurement 
system featuring not only high astrometric precision, but above all 
suited to good calibration, both from sky measurement and metrology. 

The most relevant specifications for the instrument outlined in this study are 
summarized in Tab. \ref{tab:params}, and further expanded in the following 
sections. 

\begin{table}
\begin{minipage}{0.48\textwidth}
    \centering
    \caption{Main design parameters \label{tab:params}}
    \vspace{2mm}
\begin{tabular}{ll}
Telescope diameter & $1\, m$ \\ 
Focal length & $EFL = 15\, m$ \\ 
Spectral range & Visible, $\lambda = 0.55 \, \mu m$ \\ 
Spectral band & $\Delta \lambda = 200 \, nm$ \\ 
Field of view & Annular \\ 
FOV radius & $\theta \simeq 1^\circ$ \\ 
Detector & sCMOS or CCD \\ 
Pixel size & $4\,\mu m$ \\ 
\end{tabular} 
\end{minipage}
\hfillx
\begin{minipage}{0.48\textwidth}
    \centering
    \caption{Optical design parameters \label{tab:OptDesign}}
    \vspace{2mm}
\begin{tabular}{l|r|r|r}
Mirror & Curvature & Distance & Conic \\ 
       & radius [mm] &  [mm] & constant \\ 
\hline 
M1 & 3,500.00 & 1,400.00 & -0.98 \\ 
M2 & 859.45 & 2,800.00 & -1.93 \\ 
M3 & 1,121.55 & 1,452.65 & -0.58 \\ 
\end{tabular} 
\end{minipage}
\end{table}

\subsection{Optical Design Deployment}
\label{sec:OptDeployment}  
\begin{figure}
\centering
\begin{minipage}{0.48\textwidth}
\centering
\includegraphics[width=0.95\linewidth]{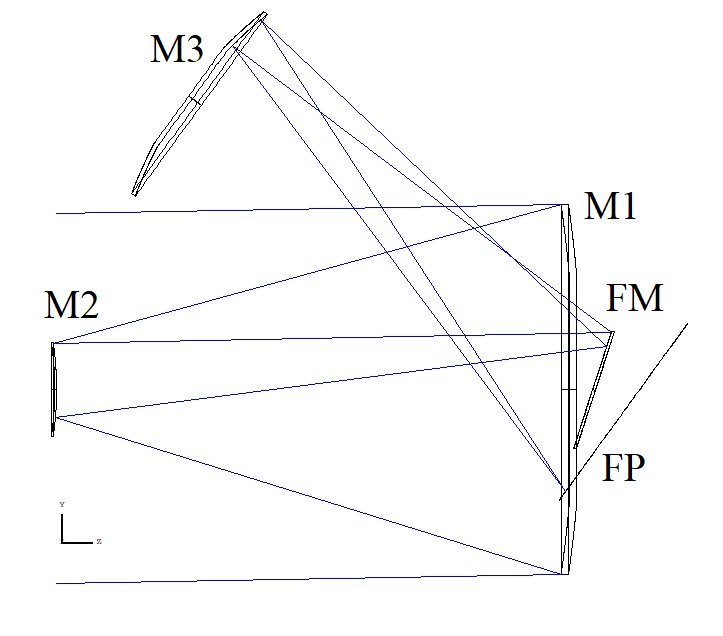} 
\caption{\label{fig:TS_2DB} Telescope layout 2D, bent configuration. }
\end{minipage}\hfill 
\begin{minipage}{0.48\textwidth}
\centering
\includegraphics[width=0.95\textwidth]{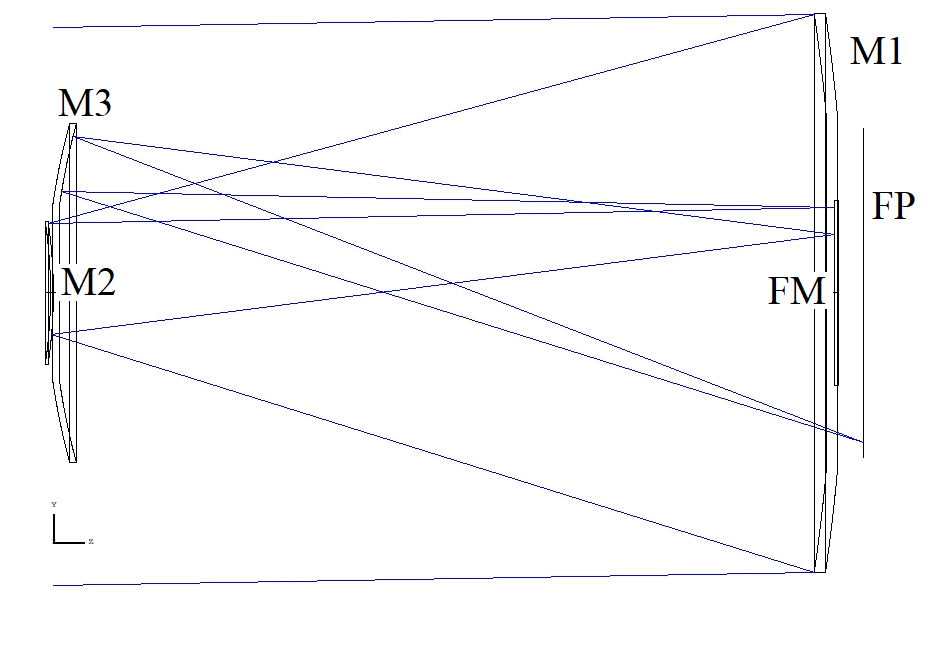} 
\caption{\label{fig:TS_2DA} Telescope layout 2D, on-axis, centered configuration. }
\end{minipage}
\end{figure}

We aim at a $1\,m$ class telescope with intrinsically good astrometric 
response on an annular field of view with $1^\circ$ radius. 
An additional asset for a space instrument is of course a compact envelope, 
which is expected to be conducive to a better trade-off between mass budget 
and mechanical stability. 
\\ 
For convenience, we adopt an example configuration from the literature 
\cite{LeeYu09}, and our optical design specifications are implemented 
by adaptation of Solution IV in that paper. 
The improvements consist in increasing the aperture and reducing the 
EFL to the desired values, and above all modifying 
the layout with the goal of a very compact configuration, as will be 
shown in Fig.\,\ref{fig:TS_2DA}. 
The optical design modelling, ray tracing and 
optimization is performed with Zemax. 
The main optical parameters are summarized in Tab.  \ref{tab:OptDesign}. 

A TMA often folds the configuration over two axes by introducing a flat 
mirror between secondary and tertiary (curved) mirrors, in order to deploy 
the optical components in separated, easily accessible regions, usually 
``behind" the primary mirror. 
Our proposed design folds the optical axis back onto itself, in order to 
minimize 
the overall instrument volume; as this only requires to turn a (flat) 
folding mirror, it is trivial from the standpoint of optical design. 
However, this allows trade-offs among different geometry constraints 
(related to vignetting and mutual obscuration), which are nonetheless 
compatible with our goal of an annular field of view.

In Figs.\,\ref{fig:TS_2DB} and \ref{fig:TS_2DA}, two layout options of 
the same underlying optical design 
are proposed; they are equivalent, since curved mirrors 
have the same shape and separation, at first order, but they are placed 
in different positions. 
On the left panel, the tertiary mirror is placed on the side of the 
input beam, whereas on the right panel it is located on-axis, close to 
the secondary mirror. 
In the former case the optical path is easier to follow, whereas the 
latter is more compact, and it is our preferred option. 
Their relative merits will be discussed in Sec.\,\ref{sec:Discussion}. 

In either figure, 
the converging beam from the primary mirror (M1, right end side) reaches 
the secondary mirror (M2, left end side), then proceeds to the flat 
folding mirror (FM) located close to the central area of M1. 
The beam is then reflected back to the tertiary mirror (M3), located 
in proximity of M2, which focuses it onto the ring of detectors of the 
focal plane (FP), again placed approximately in the intermediate 
region between FM and M1. 

\begin{figure}
	\centering
	\begin{minipage}{0.49\textwidth}
		\centering
		\includegraphics[width=0.98\textwidth]{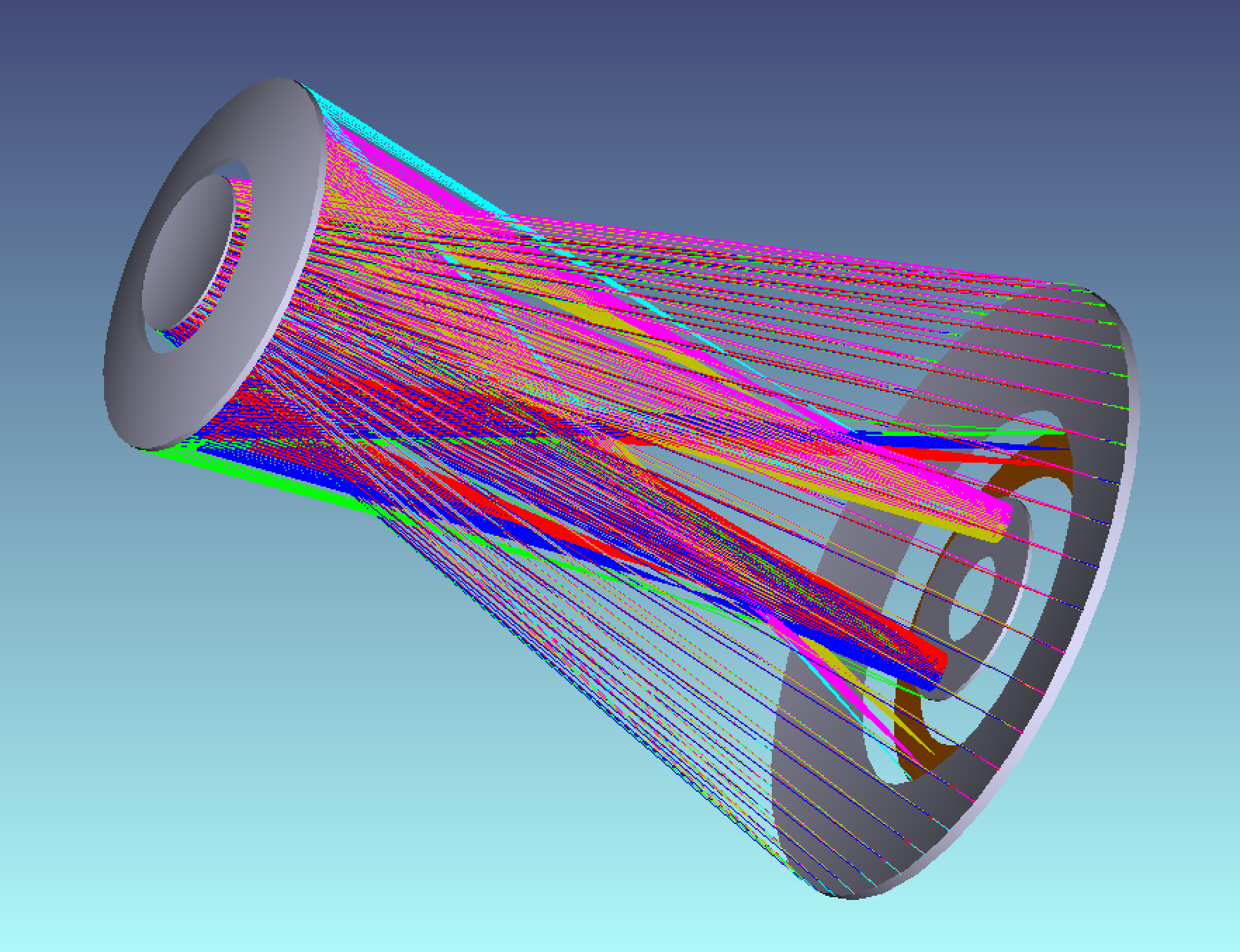} 
		\caption{\label{fig:TS_3D} Telescope 3D view, layout A: on-axis configuration } 
	\end{minipage}\hfill 
	\begin{minipage}{0.48\textwidth}
		\centering
		\includegraphics[width=0.95\textwidth]{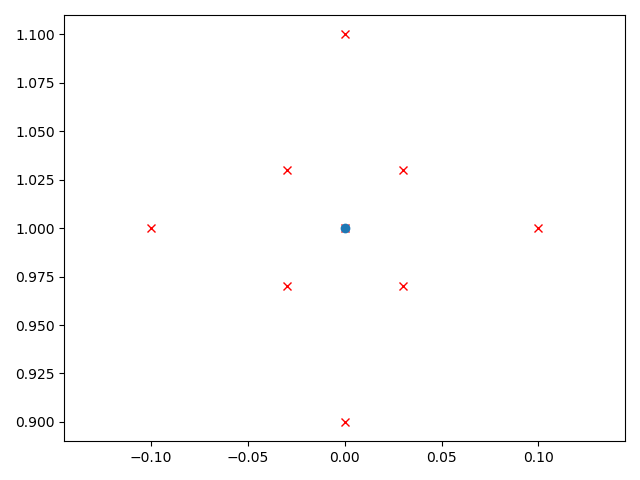} 
		\caption{\label{fig:LocField} Field positions selected for analysis around 
			the reference position $(0, 1^\circ)$. }
	\end{minipage}
\end{figure}

A three-dimensional representation of the instrument is shown in 
Fig.\,\ref{fig:TS_3D}. 
Due to the annular field, with the selected parameters, no mutual 
vignetting exists among M1, M2, FM and M3. 



\subsection{ Optical performance}
\label{sec:OptPerf}
The optical performance results in diffraction limited 
imaging over the whole annular region between radial 
values $0.9^\circ$ and $1.1^\circ$, 
so that the corrected field exceeds 
$\sim 2 \pi \times 0.2 = 1.256$ square degrees. 

\begin{sidewaystable}
    \centering
    \caption{Wavefront RMS error in selected field positions, in nm \label{tab:WFERMS}}
    \vspace{2mm}
\begin{tabularx}{0.9\linewidth}{lccccccccccc}   
Field & Total &  Distor- & Distor- & Astigm- & Astigm- &  Defocus & Coma &  Coma  & Trefoil &  Trefoil &  Residual \\ 
position  &       & tion X   & tion Y  & atism X & atism Y &          &  X   &    Y   &      X  &     Y    &  \\ 
\hline 
1: $(0, 1^\circ)$   (N)         &  11.876 &  0.000 &  1.871 &  3.052 & 0.000 & 0.109 & 0.014 & 0.080 &  0.000 & 11.140 & 2.021 \\
2: $(-1^\circ, 0)$  (E)         &  11.876 &  1.871 &  0.000 &  3.051 & 0.000 & 0.104 & 0.088 & 0.014 & 11.140 &  0.000 & 2.020 \\
3: $(1^\circ, 0)$   (W)         &  11.876 &  1.871 &  0.000 &  3.051 & 0.000 & 0.116 & 0.060 & 0.014 & 11.140 &  0.000 & 2.026 \\
4: $(0, -1^\circ)$  (S)         &  11.876 &  0.000 &  1.871 &  3.052 & 0.000 & 0.121 & 0.014 & 0.052 &  0.000 & 11.140 & 2.026 \\
5: $(0, 1^\circ.1)$             &  38.860 &  0.000 & 29.427 & 19.421 & 0.000 & 0.784 & 0.090 & 0.261 &  0.002 & 14.871 & 6.738 \\
6: $(0, 0^\circ.9)$             &  31.318 &  0.000 & 24.765 & 16.329 & 0.000 & 0.648 & 0.076 & 0.262 &  0.001 &  7.971 & 6.060 \\
7: $(0^\circ.03, 0^\circ.97)$   &  16.313 &  0.293 &  9.537 &  7.867 & 0.487 & 0.276 & 0.041 & 0.179 &  0.941 & 10.130 & 3.047 \\
8: $(-0^\circ.03, 0^\circ.97)$  &  16.313 &  0.293 &  9.537 &  7.867 & 0.487 & 0.274 & 0.032 & 0.184 &  0.942 & 10.130 & 3.047 \\
9: $(0^\circ.03, 1^\circ.03)$   &  14.400 &  0.196 &  6.766 &  2.736 & 0.159 & 0.108 & 0.009 & 0.025 &  1.065 & 12.154 & 2.277 \\
10: $(-0^\circ.03, 1^\circ.03)$ &  14.400 &  0.196 &  6.766 &  2.736 & 0.159 & 0.106 & 0.016 & 0.027 &  1.064 & 12.154 & 2.277 \\
11: $(0^\circ.1, 1^\circ)$      &  11.689 &  0.050 &  0.506 &  2.119 & 0.429 & 0.077 & 0.022 & 0.058 &  3.330 & 10.808 & 1.940 \\
12: $(-0^\circ.1, 1^\circ)$     &  11.689 &  0.050 &  0.506 &  2.119 & 0.429 & 0.077 & 0.002 & 0.062 &  3.331 & 10.808 & 1.939 \\
\end{tabularx} 
\end{sidewaystable}

\begin{figure}
\begin{center}
\includegraphics[width=0.95\linewidth]{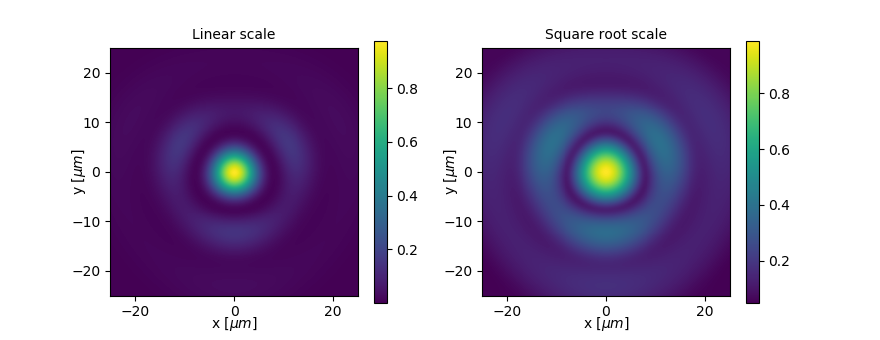} 
\end{center}
\caption{\label{fig:PSF_LinSQ} Polychromatic PSF in field 1, $1^\circ$ N, 
in linear (left) and square root (right) scale. } 
\end{figure}

The maps have a clear structure with three periods around the pupil, 
corresponding to trefoil in terms of classical aberrations 
(Zernike polynomials $n = 3,\, m=\pm 3$). 
This is actually the dominant term along the $1^\circ$ ring, 
accounting for most of of RMS WFE 
$11.14\,nm$. 
Small amounts of distortion ($1.87\,nm$) and astigmatism ($3\,nm$) 
are also present, with a residual of $2\,nm$. 
It may be noted that trefoil, given its symmetry, is expected to 
induce negligible systematic astrometric errors in the  azimuthal 
direction \cite{Busonero06}. 
The dominant trefoil results in the trilobated structure of the PSF 
(Fig.\,\ref{fig:PSF_LinSQ}), particularly evident in the first diffraction 
ring. 

In Tab. \ref{tab:WFERMS}, X and Y are angular coordinates on the tangent 
plane, corresponding respectively to radial and  azimuthal directions for 
field 1. 
The radial coordinate is reversed in field 4, and aligned with $\pm Y$ 
axis in fields 3 and 2, respectively. 
The circular symmetry is evidenced e.g. by the exchange of X and Y components 
of distortion and trefoil. 
The WFE analysis yields the same results for all fields over the $1^\circ$ 
radius circle, in the local frame of radial and  azimuthal coordinates. 
\\ 
Field positions 5 to 12, describing a small area around the reference field, 
and shown in Fig.\,\ref{fig:LocField}, 
correspond to the size of one detector (7 to 10), and to the radial size of 
the corrected field. 

\begin{figure}
\centering
\begin{minipage}{0.49\textwidth}
\centering
\includegraphics[width=0.98\textwidth]{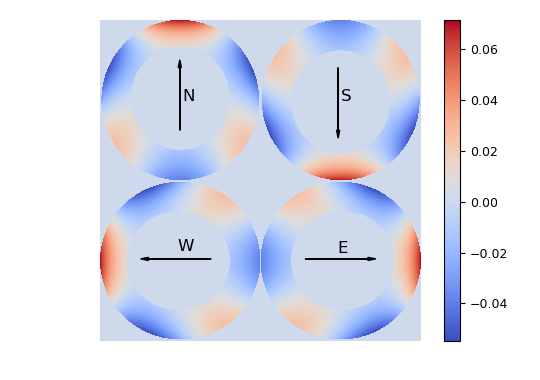} 
\caption{\label{fig:WF_maps} Wavefront error maps, in waves at $550\, nm$ 
at $1^\circ$, at four equally spaced positions (fields 1, 2, 3 and 4 in 
Tab. \ref{tab:WFERMS}) 
along the ring at $1^\circ$ from the optical axis. } 
\end{minipage}\hfill 
\begin{minipage}{0.48\textwidth}
\centering
\includegraphics[width=0.95\textwidth]{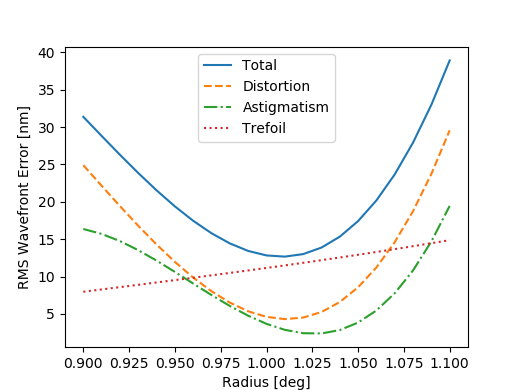} 
\caption{\label{fig:WFE_rad} RMS wavefront error and its main components 
vs. radial coordinate. }
\end{minipage}
\end{figure}

The annular region ($0.9^\circ < $ radius $ < 1.1^\circ$) is limited by 
increasing values of mainly astigmatism and distortion; actually, the latter 
does not degrade imaging quality, but only affects astrometric error, 
exclusively in the radial direction. 
Not including distortion, the RMS WFE at the ring borders is still 
$\sim 25\, nm$, i.e. better than $1 / 20$ wavelengths at 550\,nm. 

The RMS WFE and its main components (distortion, astigmatism and trefoil) 
are shown in Fig.\,\ref{fig:WFE_rad} as a function of the radial distance 
from the optical axis. 
The minimum value is close to $1^\circ$, but the aberration balancing is 
not exactly coincident for all components. 
The PSF at $1^\circ$, produced by a near-Solar unresolved source, using 
a $20\%$ spectral bandwidth (approximating a full band, filter-less 
instrument), is shown in Fig.\,\ref{fig:PSF_LinSQ} on a linear (left) and 
square root (right) scale. 
The first diffraction ring around the central lobe is remarkably brighter 
than that of a conventional filled circular aperture, due to the large 
central obstruction.

The PSF is symmetric (zero skewness) with respect to the radial axis, 
due to the underlying circular symmetry of the system; aberrations 
introduce small asymmetric contributions only along the radial direction 
(skewness: 0.2), i.e. with respect to the  azimuthal axis. 
The resulting correlation between components of the photo-centre estimate, 
according to the mathematical framework detailed in the Appendix, is 
therefore expected to be negligible. 
Numerical tests, performed in high SNR conditions, actually provide 
results for the Pearson correlation coefficient in the range $10^{-15}$ 
to $10^{-18}$.

\section{Discussion}
\label{sec:Discussion}  
Moreover, modern optical design tools are able to achieve optimal
ray tracing solutions using quite complex user-defined strategies.
It is thus possible to require good performance over the selected
ring of the focal plane, even at the expense of penalties on image
quality at the centre of the field. 
This contrasts with ``old fashioned''
analytical approaches pre-dating recent developments in computing
power, in the epoch (very few decades ago!) mainly based on perturbation
theory. In such context, on-axis performance was typically the main
criterion for the initial design step, then third order aberrations
were considered, and successively higher terms were introduced (fifth
order), in an attempt to achieve satisfactory optical performance
by aberration balancing. 
The technique is not superseded by modern ray tracing,
which relies heavily on it, and just takes advantage of numerical
``brute force'' to explore a much larger parameter space than suited
to direct human investigation. 

Conventional optical designs based on the requirement of a large field
of view usually allocate it to a large contiguous region (square,
circular, hexagonal or other) around the optical axis. The ray tracing
optimization process achieves sufficient image quality over the overall
field, usually with degradation increasing with the distance from
the axis itself, where usually the response is (nearly) diffraction
limited. Spectroscopic instruments may not have actual image quality
requirements, but their performance is nonetheless based on limited
wavefront error (WFE) over the field, e.g. to ensure adequate matching
of point-like sources to the slit or fiber feeding the dispersing system. 

The FOV requirement is sometimes actually related to a contiguous
area, e.g. for imaging large sky regions, e.g. Solar System planets,
the Moon, globular clusters and so on, but it may be relaxed if the
science needs can be expressed just in terms of achieving simultaneous
measurement of a sufficiently large area to ensure adequate source
statistics, e.g. availability of many stars above a given magnitude
threshold. In particular, exo-planet astrometry requires determination
of the distance between a bright target with respect to a reference
source of comparable magnitude, or several sources not much fainter.
Since the number of bright stars is limited, a fairly large accessible
FOV is considered as a crucial aspect. 

\subsection{Exploitation of image symmetry 
\label{sec:LocPrec}}
The concept of symmetry is intrinsically appealing, but it is necessary 
to identify practical tools for assessment of the degree of success 
toward our goals. 
Every detector of the focal plane covers the same small range of 
angular positions with respect to the optical axis, since they are all 
interchangeable by a simple rotation of the overall system. 
This is a crucial element of the planned calibration principle. 
Also, in each device, the main directions of the pixel array (rows and 
columns) can in principle be aligned with local radial and  azimuthal 
axes. 

\begin{table}
    \centering
    \caption{ Astrometric error vs. azimuthal field position, using Maximum 
    Likelihood estimator and the central PSF as reference, in the radial 
    and azimuthal direction, either ignoring or taking into account the local 
    field rotation. 
    \label{tab:AstromErr}}
    \vspace{2mm}
\begin{tabularx}{0.55\linewidth}{lccc}   
Field & Radial &  Azimuthal & Azimuthal \\ 
position &       & Non Rotated   & Rotated \\ 
      & [$mas$]  & [$\mu as$] & [$\mu as$] \\ 
\hline 
1: $(0, 1^\circ)$   (N)         &  0.0000 & 0.0000 & 0.0000 \\
2: $(-1^\circ, 0)$  (E)         &  0.0004992 & 3.592e-12 & 0.0000 \\
3: $(1^\circ, 0)$   (W)         &  -0.01504 & 4.131e-12 & 0.0000 \\
4: $(0, -1^\circ)$  (S)         &  -0.01554 & -3.709e-12 & 0.0000 \\
5: $(0, 1^\circ.1)$             &  -0.8184 & -5.325e-12 &  -5.325e-12 \\
6: $(0, 0^\circ.9)$             &   0.7836 & -3.853e-12 &  -3.853e-12 \\
7: $(0^\circ.03, 0^\circ.97)$   &   0.2198 & -39.72     &  -0.0028 \\
8: $(-0^\circ.03, 0^\circ.97)$  &   0.2198 &  39.72     &   0.0003 \\
9: $(0^\circ.03, 1^\circ.03)$   &  -0.2279 &  -16.3     &   0.0015 \\
10: $(-0^\circ.03, 1^\circ.03)$ &  -0.2279 &   16.3     &   0.0001 \\
11: $(0^\circ.1, 1^\circ)$      &  -0.0125 & -82.85     &   0.0010 \\
12: $(-0^\circ.1, 1^\circ)$     &  -0.0125 &  82.85     &  -0.0024 \\
\end{tabularx} 
\end{table}

In the nominal system, residual aberrations can be expected to be characterized 
by well defined symmetry; most of them, e.g. spherical, astigmatism, trefoil, 
distortion and coma, are symmetric in the  azimuthal direction, although some 
are not along the radial direction. 
Therefore, the PSF is symmetric in the  azimuthal direction, and at most it 
can have asymmetric terms in the radial direction. 
As derived in the appendix, this means that the photo-centre estimates in the two 
directions is not correlated, and this is a crucial property with respect to 
control of systematic astrometric errors. 
\\ 
It can be expected that, within the limit of small perturbations, the image symmetry 
properties will be retained, provided the real systems are not exceedingly removed 
from the nominal symmetry conditions. 
The issue can be addressed with standard tools of advanced optical design, 
in particular tolerancing analysis. 

\begin{figure}
\begin{center}
\includegraphics[width=0.65\linewidth]{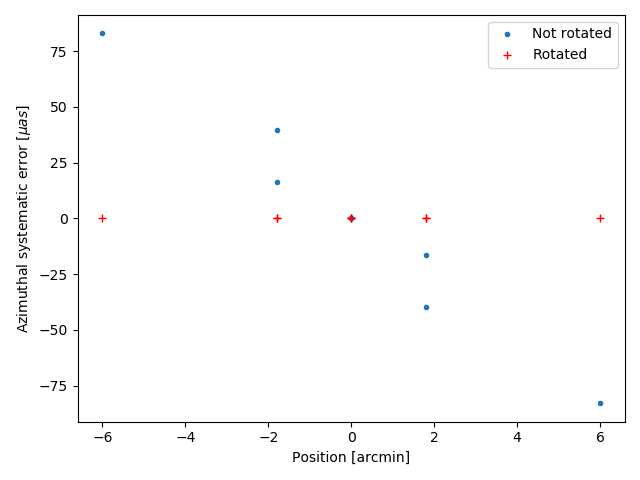} 
\end{center}
\caption{\label{fig:AstromErr} Astrometric error vs. azimuthal field position, 
using Maximum Likelihood estimator and the central PSF as reference, in the radial 
and azimuthal direction; dots: field rotation not considered; crosses: field 
rotation included. } 
\end{figure}

For the set of field positions considered in Tab. \ref{tab:WFERMS}, the 
reconstructed PSF photo-centre is computed, using a Maximum Likelihood 
estimator\cite{Mendez2013, Gai2017PASP} and the central position PSF 
(coordinates $(0,\,1^\circ)$) as reference. 
The systematic error, i.e. the PSF displacement with respect to the template, 
is listed in Tab. \ref{tab:AstromErr}, and (for field positions 5 to 12) 
shown in Fig.\,\ref{fig:AstromErr}. 
We remark that values are reported in $mas$ for the radial coordinate, and in 
$\mu as$ for the azimuthal coordinate. 

Field positions 2, 3 and 4 provide displacement close to zero with respect to 
the template, apart the natural rotation of the local reference system, 
due to symmetry reasons. 
The discrepancy with respect to mathematical zero is considered to be 
originated by sampling and numerical approximations; it is larger (but 
still a small fraction of $1\,mas$!) in the radial direction, which does not 
benefit from symmetry, and much smaller in the azimuthal direction. 

In a local region around the reference N position (fields 5 to 12), the location 
estimate provides radial errors below $1\,mas$, and azimuthal errors within 
a few $10\,\mu as$, if the local rotation of the PSF is not taken into account; 
when properly including such geometric factor, the azimuthal error drops to the few nano-arcsec range (last column of Tab. \ref{tab:AstromErr}), which appears 
consistent with the model limitations. 

{\em 
Therefore, the nominal RAFTER configuration appears particularly well behaved 
with respect to robust astrometric measurement, above all along the azimuthal 
direction, over the whole 1.256 square degrees corrected field. 
}
\subsection{ Engineering aspects } 
The telescope layout depicted in Figs.\,\ref{fig:TS_2DB}  and \ref{fig:TS_2DA} 
is proposed in two options. 
In the former, the optical path is bent to get out M3 of the input beam. 
In the latter, the folding mirror reflects the optical axis back 
onto itself, placing the tertiary mirror M3 close to the secondary 
mirror M2. 
Although in the former (labelled ``B" for ``Bent") the beam path is easier 
to follow, we reckon that the latter (labelled ``A" for ``on-Axis") 
is worth taking into account because of its compactness. 
The drawback of layout A consists of course in the large central 
obscuration of the beam, traded off in exchange for the compact layout. 
The rationale is not only associated to efficient usage of the precious 
volume of a space mission payload, but rather in the expected advantages 
in terms of higher stiffness and/or lower mass budget of the system. 

From the strictly optical standpoint, layout A suffers from lower throughput 
than B, because the central obscuration increases from $\sim 30\%$ to 
$\sim 60\%$ in linear terms, resulting in a collecting area penalty of 
$\sim 9\%$ and $\sim 36\%$ respectively. 
This holds under the assumption that the primary mirror diameter is 
retained to the same value ($1\,m$ in our case). 
However, this is not necessarily the only way to compare the two options. 
If we set constraints on the overall volume or mass budget, rather than 
on M1 diameter, the balance shifts, and it is less obviously in favour 
to option B. 
\\ 
In particular, with fixed total mass, layout A is compatible with a larger 
M1 diameter (M1 is intrinsically lightweighted by the larger central 
obstruction), thus recovering at least part of the collecting area, and 
improving on the imaging resolution. 
If we set equal M1 area, i.e. matching photon budget, layout A 
requires a primary diameter $D_1 = 1.2\,m$, and provides $20\%$ 
better resolution, with marginal changes in the remaining parts 
of the optical system. 
An equal astrometric performance criterion would result in a M1 
diameter $D_1 \simeq 1.1\,m$. 

The on-axis layout A benefits from a further asset, due to the 
naturally grouping of optical components in two subsets, including 
respectively M2 and M3 (left side in figure), and M1, FM and FP 
(right side), with a significant separation between them. 
The two groups of component are faced to each other, and might in principle 
be connected by a stable structure (e.g. a cylindrical Serrurier-like truss). 

It is possible to envisage the manufacturing of M2 and M3 from a single 
blank, e.g. by diamond turning, or their mounting on a common bench, e.g. 
by optical contact bonding. 
This would ensure a very stable geometry of the subsystem, established 
at manufacturing stage; the actual optical alignment between M2 and M3 then 
depends exclusively on FM, as its tilt may perturb, or correct, 
the coincidence between individual optical axes. 
\\ 
The M1, FM and FP subset can also be mounted on an optical bench (e.g. 
a Gaia-like torus, but much smaller), to achieve good mutual 
stability. 
Thus, the optical configuration would be preserved by mutual alignment 
of the two sub-systems, faced to each other. This does not necessarily 
alleviates all mechanical stability requirements, but at least it 
provides a simple geometric framework. 

In layout B (Fig.\,\ref{fig:TS_2DB}, lower right), some of the sensors 
are placed closer to M1 and FM than others, and could be suppressed if 
necessary, whereas layout A benefits of a more uniform 
geometry, which is expected to be beneficial to the overall 
system symmetry and minimization of systematic errors. 

\begin{table}
    \centering
    \caption{ Detailed report of the Zemax sensitivity analysis focused on the most 
     critical parameters (``offenders''). 
    Analysis conducted for a representative wavelength of 632.8 nm. 
    Optical elements labels are reported in Figure \ref{fig:TS_2DA}. 
    Z axis is the optical axis. 
    All parameters not listed give negligible contribution.
    \label{tab:SensitiviyZemax}}
    \vspace{2mm}
\begin{tabularx}{0.5\linewidth}{cccc}   
Element & Parameter & Variation & Change  \\ 
\hline 
M1       &  Tilt X and Y & +/- 3' & 21.44 \\
M2       &  Tilt X and Y & +/- 3' & 5.48 \\
M1       &  Movement X and Y & +/- 0.2 $\mu m$ & 2.27 \\
M2       &  Movement X and Y & +/- 0.2 $\mu m$ & 2.24 \\
FM       &  Tilt X and Y & +/- 3' & 1.97 \\
M1 - M2  &  Distance & +/- 0.2 $\mu m$ & 1.78 \\
M3       &  Tilt X and Y & +/- 3' & 0.47 \\
M1       &  Surface irregularity & + 0.2 Fringes & 0.46 \\
M2       &  Surface irregularity & - 0.2 Fringes & 0.22 \\
M1       &  Radius & - 1 Fringe & 0.13 \\
M2       &  Radius & + 1 Fringe & 0.09 \\
M2 - FM  &  Distance & + 0.2 $\mu m$ & 0.04 \\
M3       &  Movement X and Y & +/- 0.2 $\mu m$ & 0.02 \\
\end{tabularx} 
\end{table}

The configuration optimization is anyway left to future, more 
detailed development phases. 
In particular, aberration balancing (Fig.\,\ref{fig:WFE_rad}) may be improved, 
especially on distortion. 
However, the corrected field already exceeds 
$\sim 2 \pi \times 0.2 = 1.256$ square degrees. 
Since a much smaller area (0.258\,square degs) in the midst of the ring field 
is actually populated with detectors, on account of cost and complexity, 
there are significant margins against perturbations degrading the optical 
quality, e.g. misalignment, since external areas are expected to be more 
sensitive to them. 

\subsection{Sensitivity to misalignment}
\label{sec:Sensitivity} 
%
Conventional sensitivity analysis is usually aimed at finding the linear 
and angular ranges beyond which the telescope imaging quality (or other 
suitable descriptors) suffers from excessive degradation. 
It can be shown that the proposed configuration is, in this respect, quite 
comparable to other TMAs with similar design parameters (aperture diameter, 
effective focal length, etc.), while providing, as discussed above, potential 
benefits toward ease of implementation of a compact, stable mechanical 
assembly. 

Tab. \ref{tab:SensitiviyZemax} reports the results of the preliminary sensitivity analysis conducted with the traditional techniques. 
We analyzed the position of the elements and their tilting, as well as the surface irregularities. 
The system has an axial symmetry that results in an equivalence of the tilts and displacements with respect to the X and Y axis. 
The M1 and M2 tilts are the most offending parameters, while all other variations, included the separations and the surface irregularities are well within an acceptable level.

\begin{figure}
\begin{center}
\includegraphics[width=0.45\linewidth]{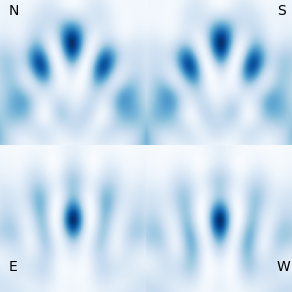} 
\includegraphics[width=0.45\linewidth]{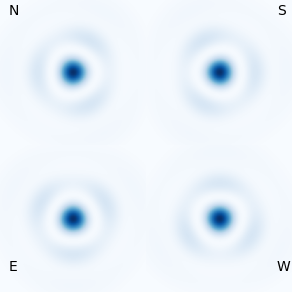} 
\end{center}
\caption{\label{fig:TiltPSF} PSF degradation for large ($36\,arcsec$, left) and 
medium ($1\,arcsec$, right) tilt of M1. } 
\end{figure}

Limiting values for M1 tilt are in the few ten arcsec range, depending on 
the actual threshold on the figure of merit, while all other components are 
retained in the nominal position; an example of PSF generated in field positions 
1 to 4 (labelled respectively N, E, W and S) of Tab. \ref{tab:WFERMS} is 
shown in Fig.\,\ref{fig:TiltPSF} (left), 
evidencing a strong degradation of the diffraction limited image, for 
a $36\,arcsec$ tilt of M1. 
Smaller perturbations generate images very similar to the nominal case, 
as shown in Fig.\,\ref{fig:TiltPSF} (right) for a $1\,arcsec$ tilt of M1. 
The M1 tilt direction is N-S. vertical in figure. 

\begin{figure}
\begin{center}
\includegraphics[width=0.32\linewidth]{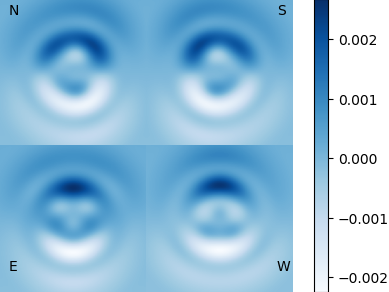} 
\includegraphics[width=0.32\linewidth]{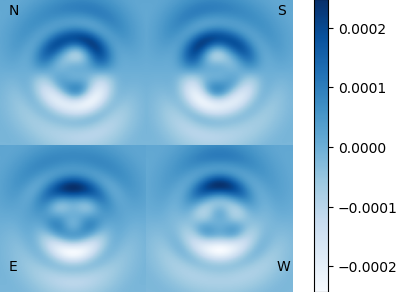} 
\includegraphics[width=0.32\linewidth]{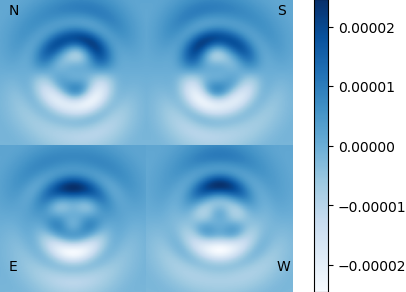} 
\end{center}
\caption{\label{fig:dPSF} PSF variation for medium (1\,arcsec, left),  
small (100\,mas, center) and tiny (10\,mas, right) tilt of M1. } 
\end{figure}

Actually, the effect is more clearly evidenced by the image difference 
with respect to the nominal PSF in the same field position, as shown in 
Fig.\,\ref{fig:dPSF} for a tilt of M1 by $1\,arcsec$ (left), $100\,mas$ (centre) 
and $10\,mas$ (right), respectively. 
As it might be expected, the graphical aspect of the PSF shape change is similar, and 
corresponding to what it could be expected from a tilt, i.e. image displacement, but 
the amplitude is clearly different, scaling approximately as the applied tilt. 
The PSF variation is quite similar in the four field positions, carrying the 
hallmark of the applied tilt, and therefore it introduces perturbations to the 
rotational PSF symmetry over the field with a clear azimuthal dependence. 
Since the PSF is computed around a nominal field position corresponding to the 
gnomonic projection of the angular position in the object space, the net effect 
appears to be an ``excess tilt''. 

The annular field of RAFTER provides therefore a convenient leverage toward 
identification of aberrations, thanks to their field dependence, contrarily 
to cases discussed in the literature according to which TMAs are difficult to 
align, because the situation taken into account was based on a single sensor 
(e.g. a WFS) in a fixed field position. 

\begin{figure}
\begin{center}
\includegraphics[width=0.42\linewidth]{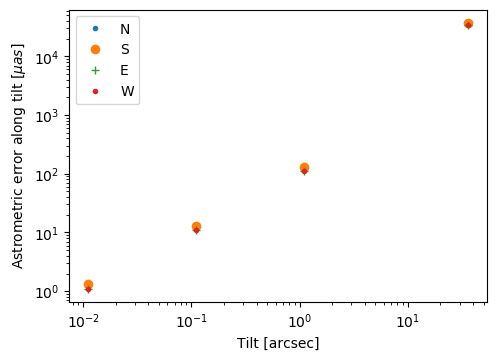} 
\includegraphics[width=0.42\linewidth]{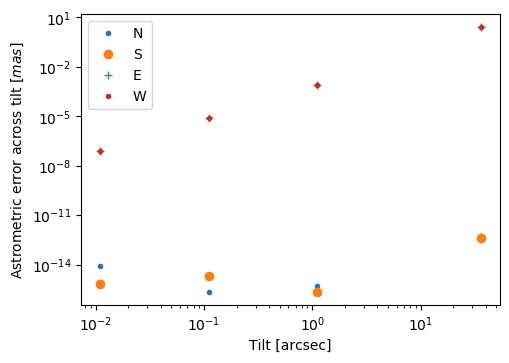} 
\end{center}
\caption{\label{fig:AstromErrTilt} Astrometric error originated by a M1 tilt in field 
    positions 1 to 4, respectively along (left) and across (right) the tilt direction. } 
\end{figure}

By application of the Maximum Likelihood estimator, and the local nominal PSF 
as reference, we compute the photo-center displacement induced by the M1 tilt. 
The results are listed in Tab. \ref{tab:AstromErrTilt}, and shown in 
Fig.\,\ref{fig:AstromErrTilt}, respectively for the direction aligned with the 
applied tilt (left panel), and orthogonally to it (right panel). 
Notably, the tilt direction is radial for fields N and S, and azimuthal for 
fields E and W. 

\begin{table}
    \centering
    \caption{ Astrometric error, in $mas$, originated by a M1 tilt in field 
    positions 1 to 4, labelled N, S, E and W. 
    \label{tab:AstromErrTilt}}
    \vspace{2mm}
\begin{tabularx}{0.7\linewidth}{lcccc}   
M1 tilt [deg] & $3e-6$ & $3e-5$ & $3e-4$ & $1e.3$ \\ 
N, radial    & -0.001289 & -0.01288 & -0.1283 & 37.4 \\ 
N, azimuthal & 8.223e-15 & -2.366e-16 & 5.236e-16 & -5.345e-13 \\ 
S, radial    & -0.001302 & -0.01303 & -0.1311 &  36.67 \\ 
S, azimuthal & -7.162e-16 & 2.035e-15 & 2.262e-16 & -3.973e-13 \\ 
E, radial    & -7.825e-08 & -7.643e-06 & -0.0007641 & -2.543 \\ 
E, azimuthal & -0.001097 & -0.01097 & -0.1098 & 34.99 \\ 
W, radial    & 7.824e-08 & 7.641e-06 & 0.000764 & 2.544 \\ 
W, azimuthal & -0.001097 & -0.01097 & -0.1098 & 34.99 \\ 
\end{tabularx} 
\end{table}

Small perturbations, about one order of magnitude below critical values evidenced 
by the optical 
sensitivity analysis (Tab. \ref{tab:SensitiviyZemax}), are associated 
to a linear range of astrometric disturbances, in this case mostly aligned 
with the tilt direction. 
The astrometric sensitivity to M1 tilt is $\sim 0.1\,mas/arcsec$, i.e. the effect 
is reduced by approximately four orders of magnitude. 
Across-tilt effects are much smaller. 
Moreover, the astrometric error associated to M1 tilt is common mode to the whole 
annular focal plane, so that it is compensated to a high degree in differential 
measurements between sources along the ring: the focal plane separation is mostly 
insensitive to the instrument perturbation. 

\section{CONCLUSIONS}
\label{sec:conclusions}  
We propose an innovative solution to the challenging problem of achieving 
a highly uniform optical response over a large field of view. 
The design driver is the achievement of an annular field of view around 
the optical axis of the telescope, with symmetry preserved to a large extent. 
The proposed design is a very compact, $1\,m$ class telescope compatible with 
modern CCD and CMOS detectors, with effective focal length EFL = $15\,m$, based 
on the established TMA telescope concept. 
We describe the concept implementation guidelines and the optical performance of 
the current configuration. The layout is extremely compact with respect to 
comparable TMA telescopes, since it is mostly deployed along the main optical 
axis. 
Circular symmetry is preserved for each optical element and for the 
overall system. 
The proposed design is considered as representative of a family of telescopes 
with the same symmetry. 
The diffraction limited FOV exceeds 1.25 square degrees, and in the proposed 
version the detector takes the best 0.25 square deg, using 66 devices. 
The detector is positioned close to the primary mirror for compactness, and 
easily interfaced to a radiator on the external envelope of the payload. 

The design is easily scalable and suited to be tailored to the needs 
of a range of space missions based on medium size telescopes and aiming at 
photon limited astrometric precision in the few $\mu as$ range. 
An example, with two configurations featuring respectively $1\,m$ and $0.7\,m$ 
primary diameter, is provided in another contribution to these Proceedings 
for the AGP mission concept. 

The good astrometric performance of RAFTER fulfills several of the goals 
settled in the ASTRA study, also presented in these Proceedings. 
Future developments may be devoted to further improvement of the optical 
performance uniformity over larger fields, optimization of an embedded 
metrology system, and inclusion of a multiple LOS beam combiner for 
large spherical angle astrometry. 

\acknowledgments   

The INAF activity has been partially funded by a grant from the Italian Ministry of 
Foreign Affairs and International Cooperation, and by the Italian Space Agency 
(ASI) under contracts 2014-025-R.1.2015 and 2018-24-HH.0. 

\bibliographystyle{spiebib}   
\bibliography{mybibl}

\end{document}